\documentstyle[sprocl,epsfig]{article}
\scrollmode
\begin{document}
\title{QCD SUM RULES AND THE LOWEST-LYING ${\bf I = 0}$ SCALAR RESONANCE}
\author{V. Elias and A. H. Fariborz}
\address{Department of Applied Mathematics \\ The University of Western 
Ontario \\
London, Ontario  N6A 5B7, CANADA}
\author{Fang Shi and T. G. Steele}
\address{Department of Physics and Engineering Physics \\ University of 
Saskatchewan \\
Saskatoon, Saskatchewan  S7N 5C6, CANADA} 
\maketitle \abstracts{A QCD Laplace sum rule framework is employed to 
investigate the
relationship between the mass and the decay width of the sigma, the lowest-
lying isoscalar
resonance in the spin-zero meson channel.  Our analysis differs from prior 
analyses, not only by
incorporating finite-width departures from the narrow resonance 
approximation, but also through
the inclusion of direct-instanton contributions and three-loop-order 
perturbative effects.  Assuming
that the sigma is the dominant subcontinuum resonance in this channel, the 
least upper bound we
obtain for the sigma- mass is shown to increase with the sigma-width, and 
suggests that a sigma 
lighter than 800 MeV should have a total width substantially smaller than 
its 
mass.}

\section{Introduction: The $\sigma$-Meson and Chiral Symmetry Breaking}
     A light $I = 0$ scalar meson [denoted henceforth as the $\sigma$] 
with a mass 
approximately equal to twice the constituent mass
of u and d quarks is common to a number of models for the dynamical 
breakdown of SU(2)
$\times$ SU(2) chiral symmetry.  Such a particle is seen to occur within 
quark-model and QCD
adaptations of the Nambu Jona-Lasinio mechanism for chiral symmetry
breaking,\cite{NJ} as well as in models for the quark-antiquark scattering 
amplitude in an 
instanton background.\cite{CC}  A light scalar
particle with a mass around 500-600 MeV is also a familiar feature of one-
boson exchange
models of the nuclear potential.\cite{SO} More recent work \cite{SV} has 
shown that the
instanton vacuum
structure predicted for QCD necessarily contains such a meson in the 500-
600 MeV
range.

     Despite its long history in theoretical models for chiral symmetry 
breakdown,
experimental evidence for the $\sigma$ has been both equivocal and 
controversial.  
Only in the last year have several independent analyses 
\cite{MS,TMR,HSS,SI} 
of $\pi-\pi$ 
scattering data 
been shown to be consistent
with a low-mass isoscalar scalar resonance, 
providing 
sufficient
experimental evidence for the $\sigma$ to be listed as the $f_0$(400-1200) 
lowest-lying scalar resonance in the present edition of the Particle Data 
Guide.
\cite{RMB} Although recent studies clearly indicate a mass for this 
particle in 
the 500-600 MeV range,\cite{HSS,SI} controversy remains as to whether this 
resonance exhibits... 

1) ... a very broad decay width comparable to its mass,\cite{TMR,RMB} as 
anticipated 
$(\Gamma_\sigma \approx 9\Gamma_\rho/2)$ from underlying chiral
symmetry, \cite{SW}

2) ... a somewhat narrower ($\approx$ 300 MeV) width,\cite{MS,HSS,SI} as 
anticipated \cite{VAM} 
from
identifying the
$\sigma$ with the
near-Goldstone particle associated with the (partially-conserved) 
dilatation current when the QCD
coupling approaches criticality, 
or

3) ... the very narrow width anticipated from the medium-range nucleon-
nucleon interaction that
has been recently extracted via inversion potentials of phase shift
data.\cite{MSHV} 
  
In view of the importance of the $\sigma$ to our present understanding of 
chiral symmetry
breaking, as
well as the controversy and uncertainty concerning its mass, width, or 
even its
existence,\cite{IS} there
is more than ample motivation to seek a QCD-based understanding of this 
lowest-lying scalar
resonance.  
\section{Field Theoretical Contribution to ${\bf I = 0}$ Scalar
Channel Sum Rules}
     QCD Laplace sum rules are well suited for determining
properties of the lowest-lying resonance in a given channel (e.g., the 
textbook \cite{PT}
QCD sum-rule extraction of the $\rho$-meson mass and decay constant in the 
vector channel),
because
they lead to an exponential suppression of higher-mass resonances not 
already absorbed into
the QCD-continuum.\cite{SVZ} QCD sum-rule methods relate phenomenological 
hadronic
physics to
perturbative QCD and order-parameters of the QCD vacuum through comparison 
of hadronic and
field-theoretical contributions to appropriate two-current correlation
functions.\cite{SVZ}

The field theoretical contributions to the Laplace sum rules
$R_{0,1}$,
\begin{eqnarray*}
\left[ 
\begin{array}{c} 
R_0(\tau) \\  R_1(\tau) 
\end{array}
\right]
\equiv \frac{1}{\pi} \left[ \int_0^\infty ds \left[ 
\begin{array}{c}
1 \\ s
\end{array} 
\right]
Im \left[\Pi(s) \right] e^{-s\tau} \right.
\end{eqnarray*}
\begin{eqnarray}
\left. - \int _{s_0}^\infty
ds \left[
\begin{array}{c}
1 \\ s 
\end{array}
\right]
Im \left[ \Pi^{pert} (s) \right] e^{-s \tau} \right],
\end{eqnarray}
are defined to be integrals over the scalar-current correlation
function
\begin{equation}
\Pi(p^2) = i \int d^4 x \; e^{i p \cdot x} < 0 | T j(x) j(0) | 0>,
\end{equation}
\begin{equation}
j(x) \equiv \left[ m_u \bar{u}(x) u (x) + m_d \bar{d} (x) d(x)\right]
/ 2.
\end{equation}
The definition (1) substracts off the hadronic continuum of states
when $s > s_0$, which is assumed to be equivalent (modulo duality) to
the purely-perturbative contribution $\Pi^{pert}$ of QCD.  Thus $R_0$
and $R_1$ are to be identified with corresponding expressions
\begin{equation}
\left[ 
\begin{array}{c} 
R_0(\tau) \\  R_1(\tau) 
\end{array}
\right]
= \frac{1}{\pi} \int_0^{s_0} ds \left[ 
\begin{array}{c}
1 \\ s
\end{array} 
\right]
Im  \left[\Pi^{res}(s)\right] e^{-s\tau}
\end{equation}
obtained from phenomenological subcontinuum-resonance contributions
$(\Pi^{res})$ to the correlation function in the $I=0$ scalar
channel, contributions which are discussed in the section that
follows.

Noting that $R_1(\tau) = - \frac{d}{d \tau} R_0 (\tau)$ (before
introducing renormalization-group improvement), we obtain
respectively the following field-theoretical QCD-vacuum-condensate, 
direct single-instanton, and purely-perturbative QCD contributions to the
$R_0$ sum rule [we work in the global-SU(2)$_f$ limit to set $m_u =
m_d \equiv m$]:

\begin{equation}
R_0(\tau) = R_0^{cond} (\tau) + R_0^{inst} (\tau) + R_0^{pert}
(\tau),
\end{equation}

\begin{equation}
R_0^{cond} (\tau) = m^2 \left[ \frac{3}{2} < m \bar{q}q> +
\frac{1}{16 \pi} <\alpha_s G^2 > - \frac{88 \pi \tau}{27} <\alpha_s
(\bar{q} q)^2> \right]. 
\end{equation}

\begin{equation}
R_0^{inst} (\tau) = \frac{3\rho^2 m^2}{16 \pi^2 \tau^3}
e^{-\rho^2/2\tau} \left[ K_0 \left( \frac{\rho^2}{2\tau} \right) +
K_1 \left( \frac{\rho^2}{2\tau} \right) \right],
\end{equation}

\begin{eqnarray*}
R_0^{pert} (\tau) = \frac{3 m^2}{16 \pi^2 \tau^2} \left[ \left(1 -( 1 +
s_0 \tau) e^{-s_0 \tau} \right) \left( 1 + \left( \frac{\alpha_s}{\pi} 
\right) 
\frac{17}{3} + \left( \frac{ \alpha_s}{\pi} \right)^2  31.864
\right) \right. \nonumber \\
-\left( \frac{\alpha_s}{\pi} \right) \left(2 + \left(
\frac{\alpha_s}{\pi}\right) \left( \frac{95}{3} \right) \right)
\int_0^{s_0 \tau} w \;  ln(w) e^{-w} dw \nonumber \\
\end{eqnarray*}
\begin{equation}
\left. + \left( \frac{\alpha_s}{\pi} \right)^2 (4.25) \int_0^{s_0
\tau} w \left[ ln (w) \right]^2 e^{-w} dw \right].
\end{equation}

     In the QCD sum-rule approach, long-distance effects are characterized 
by QCD-vacuum
condensates as in (6), local matrix elements of quark and gluon operators 
averaged over the
physical
vacuum. The QCD vacuum condensate contributions (6), which have been
known for some time, \cite{SVZ,RYR} do not distinguish between $I =
0$ and $I = 1$ in either the scalar or pseudoscalar channels.  
However, such {\it local} condensates, corresponding to vacuum 
fluctuations with
infinite
correlation length, are insufficient in the scalar and pseudoscalar 
channels to account for the full
nonperturbative content of the QCD vacuum -- they do not take into account 
the 
{\it nonlocal}
contributions to current correlation functions arising from instanton-
induced vacuum
fluctuations.\cite{DEK}
The failure to find a light scalar resonance in previous QCD sum-rule
treatments,\cite{RYR,EVS} can be attributed in part to the failure to
incorporate an explicit (or sufficiently well-understood) direct
instanton contribution to the two-scalar-current correlation
function. 
The instanton contribution (7) to $R_0$ is the
same as the known \cite{DEK,ES} 
instanton contribution to $R_0$ in the $I = 1$
pseudoscalar channels. 
The equivalence to the $I = 0$
scalar channel follows from cancelling sign changes between the
isostructure and the $\gamma$-matrix traces in going from $I = 1$
pseudoscalar to $I = 0$ scalar channels. \cite{RDC} The parameter
$\rho$ represents the single-instanton size.   
Finally, the purely perturbative contribution(8) is obtained by
substituting the explicit three-loop perturbative QCD 
contribution to
the scalar-current correlation function, which can be extracted from
an expression calculated by Gorishny et al, \cite {SGG} directly into
eq. (1).

\section{Finite-Width Corrections to Lowest-Lying Resonance Masses}

      In the usual narrow resonance approximation, resonance contributions 
to
$Im\Pi(s)$ appearing
in (1) are proportional to $\delta$-functions at the resonance mass:
\begin{equation}
Im \Pi^{res}(s) = \sum_{r} \pi g_r \delta (s - m_r^2).
\end{equation}
It is evident from substitution of (9) into (1) that higher-mass resonance 
contributions 
to $R_k$ are exponentially suppressed:
\begin{equation}
R_k^{res}(\tau) = \sum_{r} g_r m_r^{2k} e^{-m_r^2 \tau}.
\end{equation}
Moreover, if $m_\ell$ denotes the lightest subcontinuum resonance mass,
then
\begin{equation}
R_1 (\tau) / R_0 (\tau) \geq m_\ell^2 .
\end{equation}
In standard sum rule methodology (such as in using the I = 1 vector 
channel sum rules to extract
the $\rho$ mass \cite{PT,SVZ}), one can minimize field-theoretical 
expressions for
$R_1(\tau)/R_0(\tau)$ over an appropriate
range of $\tau$ to obtain via (11) a least upper-bound on the lowest-lying 
resonance mass
$m_\ell$.

Relations (9-11) require modification to take into account nonzero 
resonance widths.  The
$\delta$-functions in (9) should be understood to be the narrow width 
limit of Breit-Wigner
resonances:
\begin{equation}
\pi g_r \delta(s - m_r^2) = \lim_{\Gamma_r \rightarrow 0} Im [-g_r /
(s - m_r^2 + im_r \Gamma_r)].
\end{equation}
The Breit-Wigner shape on the right hand side of (12) may be expressed as 
a 
Riemann sum  of unit-area pulses $P_{m_r}$ centred at $s = m_r^2$:
\cite{EFS}
\begin{equation}
P_M(s,\Gamma) \equiv \left[ \Theta(s - M^2 + M \Gamma) - \Theta(s -
M^2 - M \Gamma) \right] / 2M \Gamma .
\end{equation}
\begin{eqnarray*}
Im \left[\frac{-1}{s - M^2 + iM \Gamma} \right] =
\frac{M \Gamma}{(s - M^2)^2 + M^2 \Gamma^2}
\end{eqnarray*}
\begin{equation}
= \lim_{n \rightarrow
\infty} \frac{2}{n} \sum_{j = 1}^{n} \sqrt{\frac{n}{j - f} - 1} \; \;  P_M
(s, \; \sqrt {\frac{n}{j - f} - 1} \; \; \Gamma), \; \; [0 \leq f < 1].
\end{equation}
In an n = 4 (four-pulse) approximation with $f$ chosen to be 0.70 to 
ensure that the area under
the
four pulses is equal to $\pi$ (the area under the Breit-Wigner curve), we 
find for nonzero widths
that
the lowest-lying resonance contributions to $R_{0,1}$ are given by
\begin{equation}
[R_0 (\tau)]_\ell = g_\ell \;  e^{-m_\ell^2 \tau} W_0 (m_\ell,
\Gamma_\ell, \tau),
\end{equation}
\begin{equation}
[R_1 (\tau)]_\ell = g_\ell \; m_\ell^2 \; e^{-m_\ell^2 \tau} W_1 (m_\ell,
\Gamma_\ell, \tau_\ell),
\end{equation}
where
\begin{eqnarray}
W_k [M, \Gamma, \tau] = 0.5589 \Delta_k (M, 3.5119 \Gamma,
\tau) + 0.2294 \Delta_k (M, 1.4412 \Gamma, \tau) \nonumber \\
+ 0.1368 \Delta_k (M, 0.8597 \Gamma, \tau) + 0.0733
\Delta_k (M, 0.4606 \Gamma, \tau)
\end{eqnarray}
\begin{equation}
\Delta_k (M, \Gamma, \tau) M^{2 k}e^{-M^2 \tau} \equiv
\int_{-\infty}^{\infty} P_M
(s, \Gamma) s^k \; e^{-s \tau} ds.
\end{equation}
Width effects are then seen to alter the expression (11) for the lowest-
lying resonance mass.  If
the lowest-lying resonance is the dominant subcontinuum resonance in a 
given channel,
we find that
\begin{equation}
m_\ell^2 \leq \left( \frac{R_1(\tau)}{R_0(\tau)} \right) \left( \frac{
W_0(m_\ell, \Gamma_\ell, \tau)}{W_1(m_\ell, \Gamma_\ell, \tau)}
\right).
\end{equation}
Thus, for a given choice of width $\Gamma_\ell$, one can use the field-
theoretical 
expressions for $R_{0,1}(\tau)$ to
obtain via (19) a self-consistent least upper bound for $m_\ell^2$.

\section{Application to the ${\bf I = 0}$ Scalar Channel}

      For a given choice of $\Gamma_\sigma, s_0$, and the Borel-parameter 
mass-scale
$M(\equiv \tau^{-1/2})$, one can use
the I = 0 scalar channel expressions for $R_{0,1}$ in conjunction with 
(19) to obtain a sum-rule
estimate of $m_\sigma$.  In Figure 1, we have displayed such estimates as 
a function of
$M$ for various
choices of $s_0 \geq 1$ GeV$^2$, assuming values for $\Gamma_\sigma$ of 
zero, 300, 400, and
500 MeV.  We have
used standard values for the parameters appearing in
(6-8): \cite{EFS} 
$<m\bar{q}q> = - f_\pi^2 m_\pi^2/4$  $(f_\pi =$ 131 MeV),
$<\alpha_s G^2>$ = 0.045 GeV$^2$, $<\alpha_s(\bar{q}q)^2>$ = 0.00018
GeV$^6$, $\rho$ = (600 MeV)$^{-1}$.  Factors of $\alpha_s$ that are not
absorbed in (approximately-) RG-invariant condensates are replaced with 3-
loop 
running ($\Lambda_{QCD}$ = 150 MeV) coupling constants $\alpha_s(M)$, 
consistent 
with $R_k$ being solutions of RG equations in the Borel
parameter $M$.\cite{NR} 
The Borel scale $M$ is itself allowed to vary over the range 0.4
GeV$^2 \leq M^2 \leq s_0$, the upper bound being a necessary condition for 
the subcontinuum 
character of the lowest-lying resonance.  A property common to all four 
graphs of Fig 1 is 
an increase in sum-rule-estimated values of $m_\sigma$ with increasing 
$s_0$.  A 
comparison of the four graphs also shows, for a
given value of $s_0$, that sum-rule estimates of $m_\sigma$ increase with 
width.  We observe
from the final
graph, for example, that the {\it global} minimum value of $m_\sigma$ is 
larger than 
800 MeV if $\Gamma$ = 500 MeV. 
If we assume, as is usual in sum-rule methodology, that a physical result 
should be locally
insensitive to the Borel parameter, then we must consider only those 
curves for which a local
minimum occurs at a value of $M$ less than $s_0^{1/2}$ [all curves 
displayed in Fig 1 are cut
off at $M^2 = s_0$].  As is evident from the Fig 1 graphs, such a {\it 
local} 
minimum does not occur unless $s_0 > 1.6$ GeV$^2$.  Moreover, the value of 
$m_\sigma$ at this local minimum increases with
increasing $\Gamma$.  In Table
I are tabulated the lowest values of the continuum threshold $s_0$ for the 
onset of a local
minimum, as well as the value of $m_\sigma$ associated with this local 
minimum
({\it i.e.}, the {\it minimum} minimizing $m_\sigma$). The results
are clearly indicative of a sigma mass that increases with width. We
have also performed a weighted least squares fit of $R_0(\tau)$, as
given in (5-8) to the $\tau$-dependence (15) anticipated for a single
lowest-lying resonance.  The fit is performed utilizing the Monte
Carlo simulation of uncertainties described in ref. 22 over the
Borel-parameter range $0.4$ GeV$^{-2}$ $\leq \tau \leq 2.2$
GeV$^{-2}$.  Results of the fit are as follows (uncertainties are
90\% confidence levels): $m_\sigma = 0.93 \pm 0.11$ GeV, $\Gamma
\leq$ 260 MeV, $s_0 = 3.15 \pm 0.96$ GeV$^2$.

As a final note, we stress that the above results are obtained by assuming
that {\it only the
lowest-lying resonance} $\sigma$ contributes to $R_{0,1}$.  While I = 0 
scalar resonances with
masses much
larger than 1 GeV may safely be regarded as exponentially suppressed (if 
not massive enough
to be absorbed in the $s > s_0$ hadronic continuum), the $f_0$(980) 
resonance must be regarded
as
subcontinuum 

{\noindent
\epsfig{file=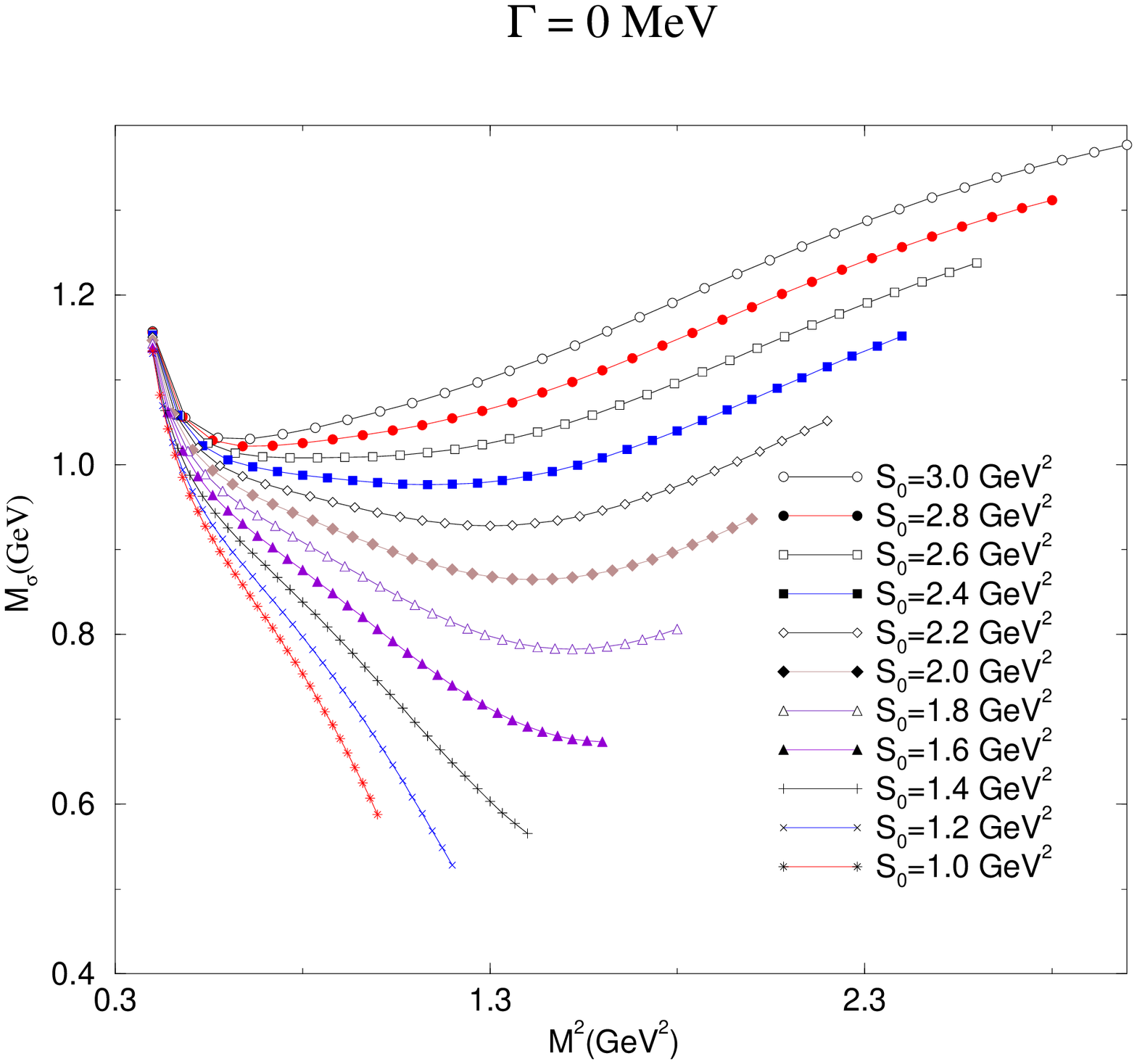,
height=2.in, angle=270}
\epsfig{file=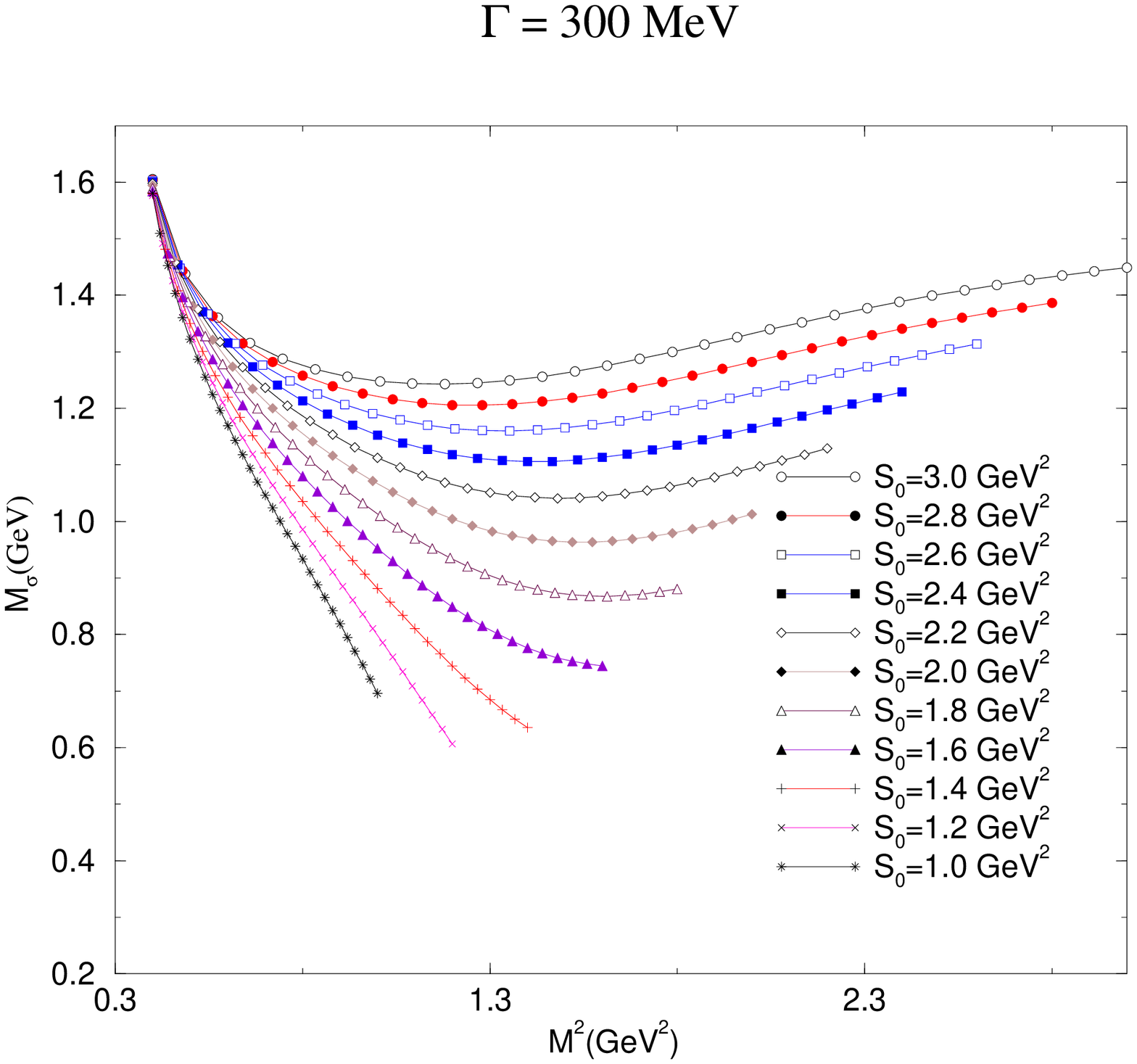,
height=2.in, angle=270}\\
\epsfig{file=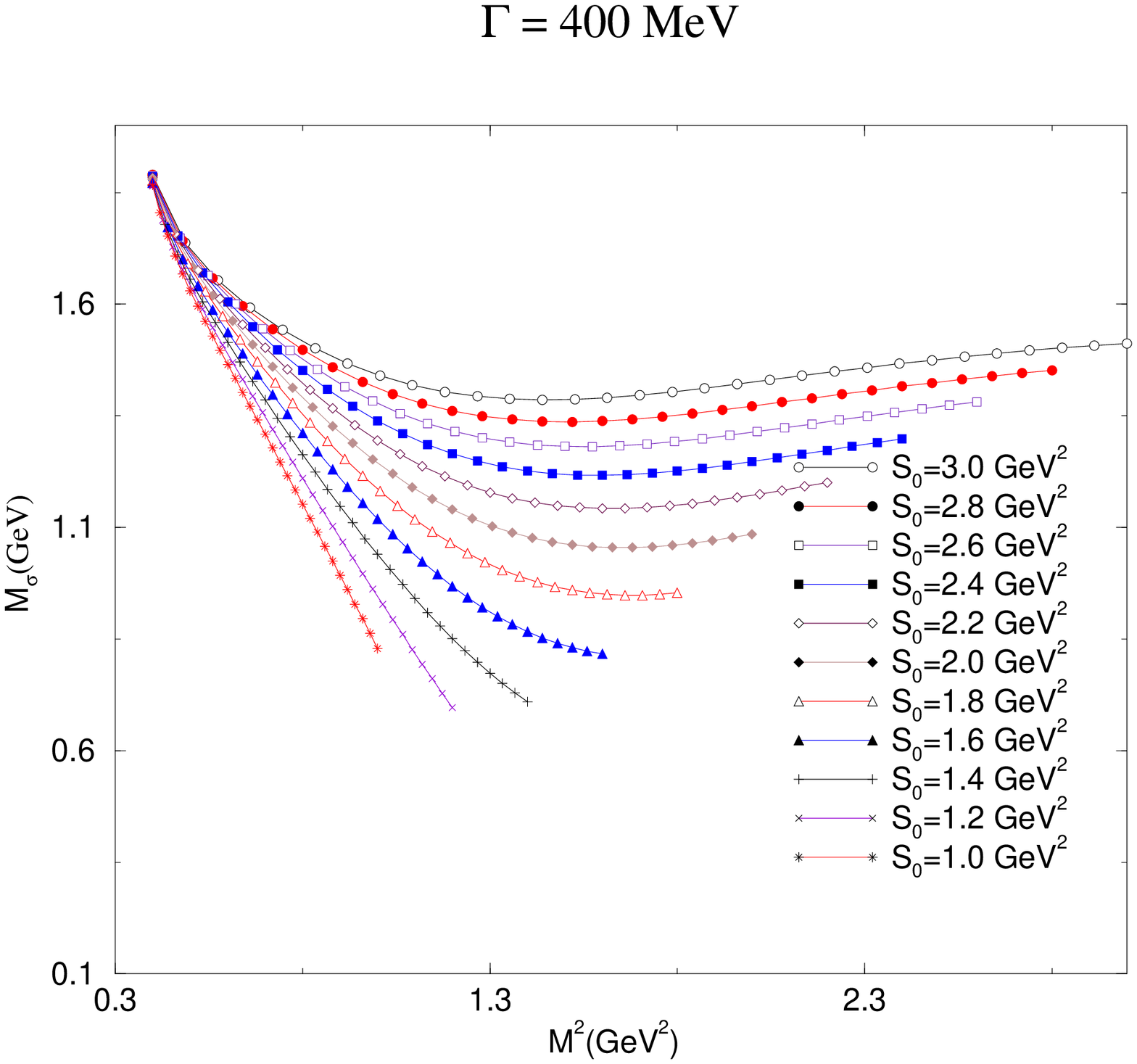,
height=2.in, angle=270}
\epsfig{file=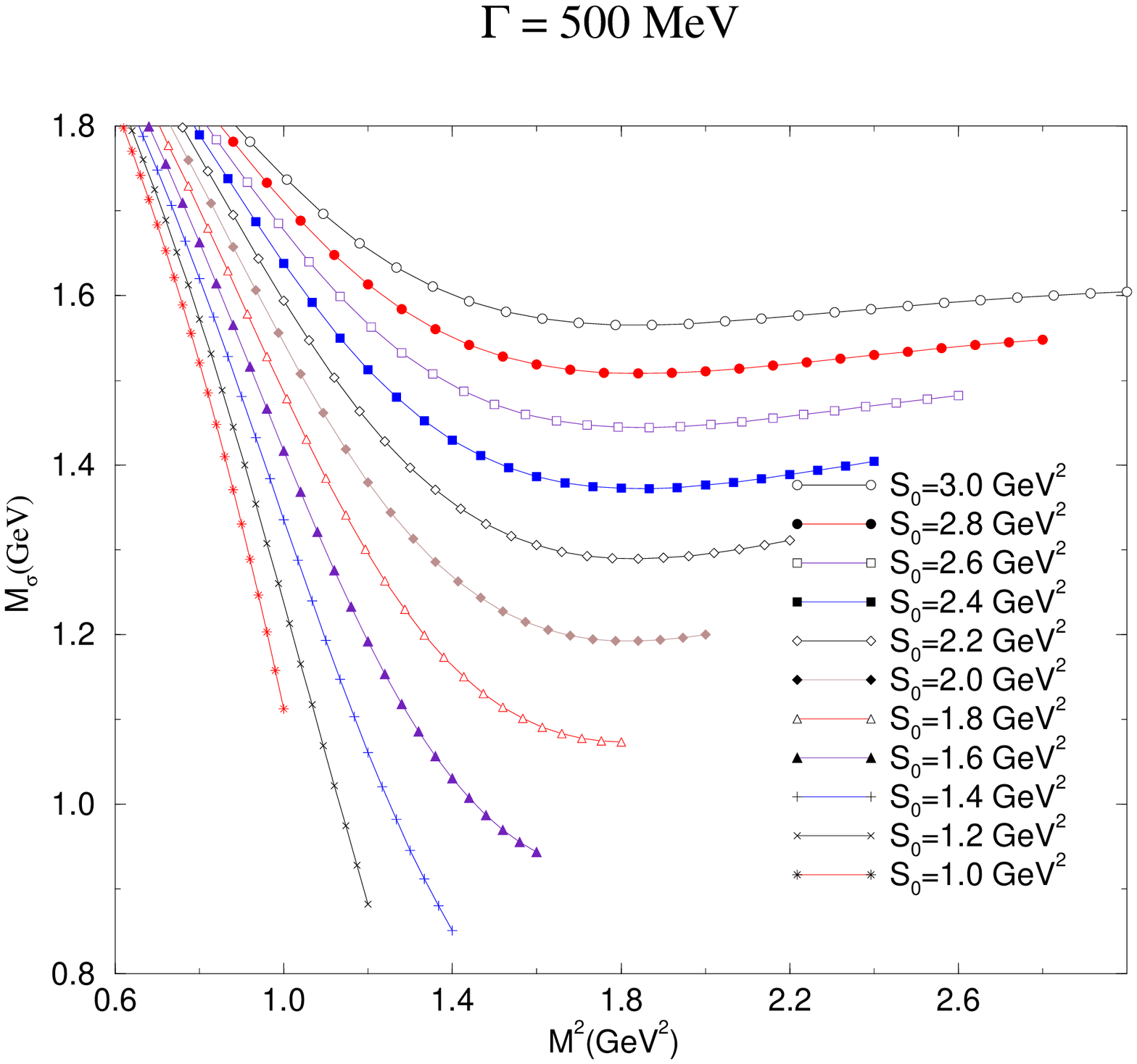,
height=2.in, angle=270}
}
\begin{center}
Fig.1: {\small Sum-rule estimates of $m_\sigma$}
\end{center}

\begin{table}[h]
\begin{center}
\begin{tabular}{|c|c|c|c|c|c|c|}
\hline
$\Gamma (MeV)$      &    0    &   100   & 200 &  300   & 400  & 500
\\ \hline
$s_0 (GeV^2)$       &  1.61   &  1.61   & 1.62  & 1.65 & 1.70 & 1.82
\\ \hline
 $m_\sigma (MeV)$   &  680    &  687    & 716   & 778  & 884  & 1087
\\ \hline
\end{tabular}
\end{center}  
\caption{Sigma mass lower-bounds associated with the onset of a local
minimum}
\label{Ms_s0_G_exact}
\end{table}  
{\noindent 
in character.  If this resonance is $s\bar{s}$, one might 
expect this resonance to
have a
Zweig-suppressed coupling to the current (3); similar suppression might be 
expected for more
exotic interpretations \cite{RMB} of the resonance. However, the 
disturbing possibility exists 
that the narrow-resonance $f_0$(980) is itself the resonance under study 
in our analysis, perhaps corresponding in the first Fig 1
graph ($\Gamma=0$) to the very flat local minimum occurring when $s_0$ = 
2.4 GeV$^2$. 
}

\end{document}